\documentclass[12pt]{article} 
\usepackage[T1]{fontenc}
\usepackage[export]{adjustbox}
\usepackage[utf8]{inputenc}
\usepackage[myheadings]{fullpage}
\usepackage[none]{hyphenat}
\usepackage[section]{placeins}
\usepackage[a4paper,margin=2.5cm]{geometry} 
\usepackage[backend=biber,style=alphabetic,citestyle=alphabetic]{biblatex}

\usepackage{algorithm,algpseudocode,amsfonts,amsmath,amssymb,amsthm,authblk,bbm,bbold,bm,caption,csquotes,enumitem,esdiff,epsfig,epstopdf,fancyhdr,graphicx,hyperref,lipsum,listings,lmodern,mathabx,mathtools,overpic,pgfplots,physics,scalerel,stackengine,stackrel,stmaryrd,subcaption,tcolorbox,tikz,url}

\usetikzlibrary{calc}
\pgfplotsset{compat=1.11}

\begin{document}
\pagenumbering{arabic}
\author[1]{Pascal Traccucci}
\author[2]{Luc Dumontier}
\author[3]{Guillaume Garchery}
\author[4]{Benjamin Jacot}
\title{A Triptych Approach for Reverse Stress Testing of Complex Portfolios}
\affil[1]{Global Head of Risk at la Fran\c caise}
\affil[2]{Partner and Head of Factor Investing at LFIS}
\affil[3]{Partner and Head of Quantitative Research at LFIS}
\affil[4]{Quantitative Risk Manager at la Fran\c caise}
\maketitle
The quest for diversification has led to an increasing number of complex funds with a high number of strategies and non-linear payoffs. The new generation of Alternative Risk Premia (ARP) funds are an example that has been very popular in recent years. For complex funds like these, a Reverse Stress Test (RST) is regarded by the industry and regulators as a better forward-looking risk measure than a Value-at-Risk (VaR). We present an Extended RST (ERST) triptych approach with three variables: level of plausibility, level of loss and scenario. In our approach, any two of these variables can be derived by providing the third as the input. We advocate and demonstrate that ERST is a powerful tool for both simple linear and complex portfolios and for both risk management as well as day-to-day portfolio management decisions. An updated new version of the Levenberg - Marquardt optimization algorithm is introduced to derive ERST in certain complex cases.\newline\newline
\textbf{Keywords}: Reverse Stress Test, Non-Linear Portfolio Theory, Quadratically Constrained Quadratic Program\newline\newline
\textbf{This article reflects the authors' opinion and not necessarily those of their employers.}
\section*{Introduction: The Case of Premia Portfolios}
    Academic theory has been mined to support the development of investment solutions containing an ever-increasing number of factors. Over the last decade, academics and practitioners have shown that traditional asset classes offer limited diversification, especially in market downturns. For example, it is shown in \cite{Callan} that equities (even combined with alternative assets such as infrastructure, real estate, high yield bonds, etc.) and government bonds are proxies for only two macroeconomic risk factors: GDP growth and inflation. In recent years though, investors have grown increasingly worried that these asset classes are expensive due to extraordinarily accommodative global monetary policy. In response, academics and practitioners have delved into Modern Portfolio Theory (MPT) to identify the microeconomic factors that are the backbone of ARP solutions. In \cite{Mesomeris}, the ARP 1.0 approach combines 10 to 15 different long/short portfolios capturing standard investment styles such as value, carry, momentum, low risk, or liquidity across a broader scope of traditional asset classes. For further diversification, the ARP 2.0 approach combines up to 30 strategies by including investment banking style premia, likely to use instruments with quadratic profiles. \cite{Dumontier} gives an example of such premia in the commodity market.\par
    As the number of factors in a complex portfolio increases, so does risk.\cite{Harvey} observed an exponential increase in factor discoveries since the CAPM and the market factor were identified in the 1960s. 
    The “zoo of factors”\footnote{John Cochrane of the University of Chicago coined this term in his 2011 presidential address to the American Finance Association.} now includes several hundred factors. This raises two concerns. All factors will not achieve simulated returns and there is the risk that a given "new" factor is just another expression of an existing one and likely to deliver correlated returns - see \cite{Suhonen}. This is especially dangerous as the calibration error of a portfolio’s volatility increases in line with the number of factors it includes, and soars if these factors – expected, in theory, to be uncorrelated – end up re-correlating strongly. As shown in \cite{Dumontier2}, the volatility of Equal Risk Contribution (ERC) portfolios that contain 10 and 30 theoretically uncorrelated factors, respectively, will double and more than triple, respectively, with a realized pairwise correlation of 30\%. Additionally, pairwise correlations tend to increase in line with individual volatility levels, further increasing the calibration error of the portfolio’s volatility. The resulting risk is that individual strategies deliver negative returns precisely when portfolio volatility is rising, resulting in heavy losses.\par
    Many risk management frameworks cannot properly account for non-linear profiles and assess the risk of loss associated with combining an unusually high number of strategies. Specifically, historical VaR is an instantaneous risk indicator and does not correspond to a clearly identified scenario, hence the need for complimentary stress tests. To build a stress testing tool, the dataset must be simplified and historical or pre-defined scenarios are used without quantifying their plausibility. Thus, parametric VaR imposes dependence on a model to benefit from an analysis framework in the form of a VaR and a sensitivity of this VaR to all the parameters of the model. This requires that several numerical problems be addressed, especially in case of quadratic profit and loss (P\&L). This paper presents an innovative approach: ERST, following on from the work in \cite{Mouy} and \cite{Packham}. This approach is able, with low technical costs\footnote{Using an algorithm derived from the Levenberg-Marquardt one to deal with complex problems.}, to deliver two of three parameters provided that the third is given as input. The three parameters are a scenario, a level of plausibility and a level of loss (see figure \ref{fig:STRESSTESTS}). The result is a more meaningful risk measure and one that corresponds to a clearly identified scenario.
    \tikzstyle{cloud} = [draw, rectangle, node distance=3cm, minimum height=2em]
    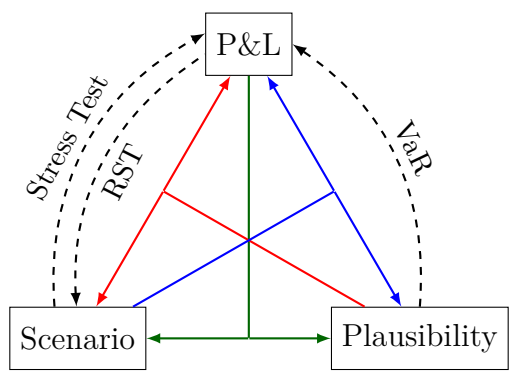
\begin{figure}
        \centering
        \begin{tikzpicture}[scale=1.5,>=latex]
            \node (x) [inner sep=0,minimum size=0] at (0,0){};
            \node (y) [inner sep=0,minimum size=0] at (-.75,1.3){};
            \node (z) [inner sep=0,minimum size=0] at (.75,1.3){};
            \node (x1) at (-1.5,0)  [cloud]{Scenario};
            \node (x2) at (1.5,0)   [cloud]{Plausibility};
            \node (x3) at (0,2.6)  [cloud]{P\&L};
            \draw[thick,->,color=black!60!green]   (x) -- (x1);
            \draw[thick,->,color=black!60!green]   (x) -- (x2);
            \draw[thick,-,color=black!60!green]    (x) -- (x3);
            \draw[thick,->,color=red] (y) -- (x1);
            \draw[thick,->,color=red] (y) -- (x3);
            \draw[thick,-,color=red]  (y) -- (x2);
            \draw[thick,->,color=blue]  (z) -- (x2);
            \draw[thick,->,color=blue]  (z) -- (x3);
            \draw[thick,-,color=blue]   (z) -- (x1);
            \draw[->, bend right=30, thick, dashed] (x2.north) to node[sloped, anchor = center, above] {\small VaR}(x3.east);
            \draw[->, bend left=33, thick, dashed] ($(x1.north west)!0.33!(x1.north east)$) to node[sloped, anchor = center, above] {\small Stress Test}($(x3.north west)!0.33!(x3.south west)$);
            \draw[<-, bend left=30, thick, dashed] ($(x1.north west)!0.49!(x1.north east)$) to node[sloped, anchor = center, below] {\small RST}($(x3.north west)!0.66!(x3.south west)$);
        \end{tikzpicture}
        \caption{The triptych approach of the Extended Reverse Stress Test (ERST) is shown in red, green and blue. Input is either a scenario, a plausibility or a P\&L and the output are the other two.
        }
        \label{fig:STRESSTESTS}
    \end{figure}
    In what follows, $\bm{S}$ is defined as a scenario. It is a vector whose length $n$ equals the number of risk factors the portfolio is exposed to. $\bm{S}$ here contains a limited number of risk factors, specifically changes in the daily returns $r_i$ or implied volatility $\delta\sigma_j$ to which the portfolio is sensitive. Other risk factors such as currency exchange rates, dividends, interest rates or repo rates and other sensitivities such as Theta can also be incorporated.\par
    In summary:
    \begin{equation*}
        \bm{S}=(\dots,r_i,\dots,\delta\sigma_j,\dots)
    \end{equation*} 
    In addition, the co-variance matrix of the risk factors will be denoted as $\mathbf{\Sigma}$.
\section{Starting from a Scenario}\label{sec:first_approach}
    A scenario-driven ERST approach is suitable for a portfolio manager considering a given adverse or best-case scenario $\bm{S}_0$. To assess the plausibility of such scenario, the probability $\alpha_0$ for a scenario equally or less extreme to $\bm{S}_0$ is computed. If $\alpha_0$ is too high, a more plausible version $\tilde{\bm{S}}$ of $\bm{S}_0$ is derived and suggested to the portfolio manager.
    \subsection{Measuring Plausibility}\label{sec:IntroMaha}
\begin{figure}
            \centering
            \includegraphics[width=.4\linewidth,trim={1cm 1.25cm 1.5cm 2.5cm},clip]{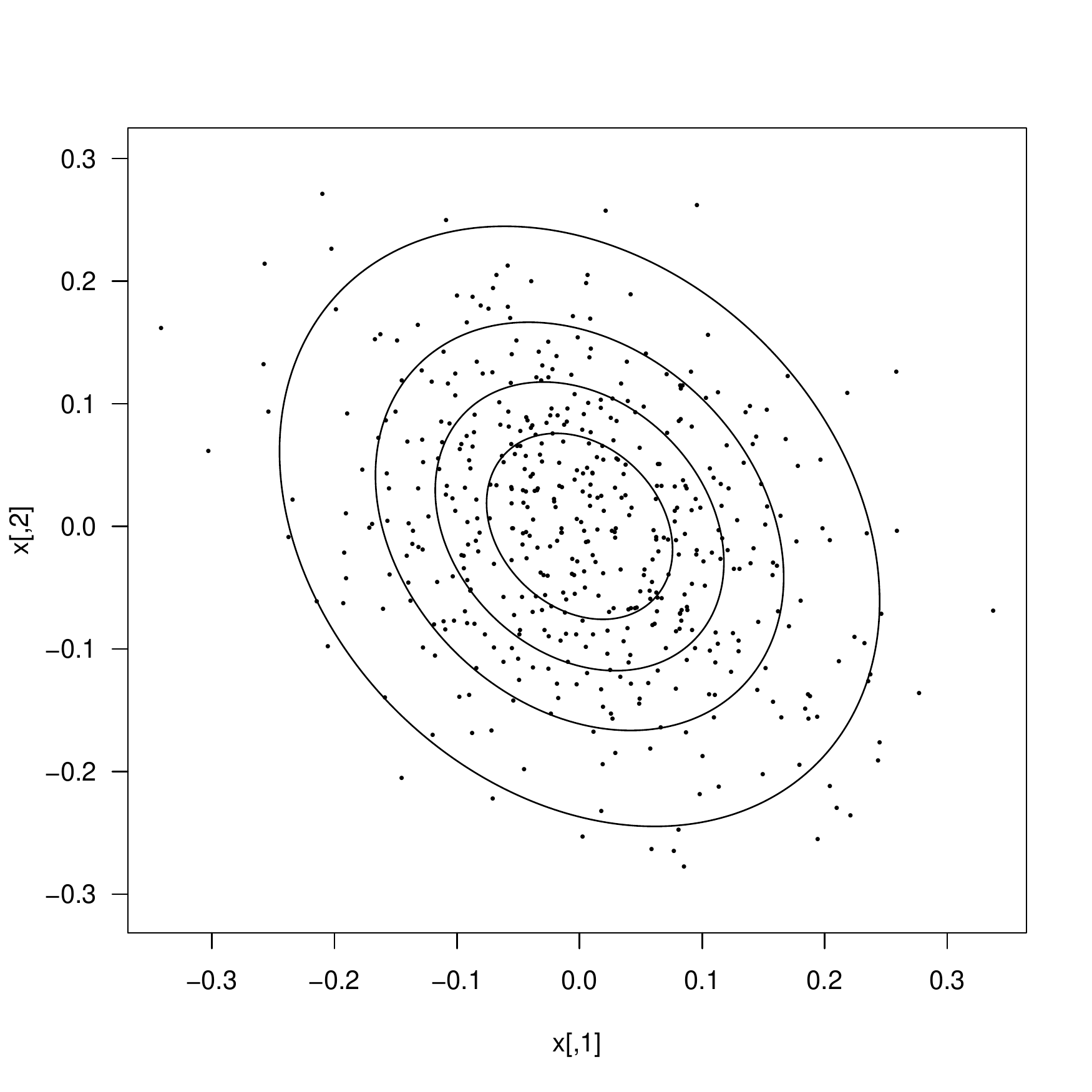}
            \caption{Plausibility domains for a bi-variate random variable with elliptical density. The inner (resp. outer) ellipse corresponds to a $25\%$ (resp. $95\%$) quantile. In-between ellipses correspond to a $50\%$ and $75\%$ quantile.}
            \label{fig:exemple1}
        \end{figure}

        ERST relies heavily on the concept of plausibility (or likelihood) to discriminate between the scenarios generated. Multiple plausibility measures exist in the literature \cite{Glasserman} \cite{Breuer_2}. In this paper, plausibility is quantified in terms of Mahalanobis distance. The latter measures the amplitude of the multivariate moves in $\bm{S}$ from the mean scenario $\bm{\mu}$ in units of standard deviation. It is therefore similar in a multidimensional space to the concept of Z-score $z$ or standardized variables. As a reminder:
        \begin{equation}
            z=\frac{x-\mu_X}{\sigma_X}
        \end{equation}
        where $x$ is a realization of a random variable $X$ with mean $\mu_X$ and standard deviation $\sigma_X$. Mahalanobis distance is defined as follows:
        \begin{equation}
            \label{eq::maha_def}
            \text{Maha}^2(\bm{S})=(\bm{S}-\bm{\mu})^\top\mathbf{\Sigma}^{-1}(\bm{S}-\bm{\mu})
        \end{equation}
        Unlike other measures, Mahalanobis distance is both intuitive and simple to use. The following characteristics of the Mahalanobis distance are noteworthy:
        \begin{enumerate}
            \item A low (resp. high) Mahalanobis distance characterizes a highly plausible (resp. unlikely) scenario.
            \item $\text{Maha}^2(\bm{S})=R^2$ is the surface of an ellipsoid of radius $R$. Points within the ellipsoid have a Mahalanobis distance of less than $R$. The further away these points are from the surface, the closer they are to the center and the more plausible they become. 
            \item  Assuming $\bm{S}$ follows a multivariate normal distribution, $\text{Maha}^2(\bm{S})$ follows a $\chi^2(n)$ distribution, as proved in \cite{Studer}. The $\alpha$ quantile of a $\chi^2(n)$ density is thus the squared radius of the ellipsoid where $\alpha\%$ of the multivariate normal scenarios $\bm{S}$ remain inside. Hereafter, this ellipsoid is referred to as $\mathcal{E}^\alpha$. See figure \ref{fig:exemple1} as an illustration.
            \item Mahalanobis distance is suited to any elliptical multivariate distribution for $\bm{S}$, which includes densities other than multivariate normal - for example, Student's t distribution. This is of primary importance as a distribution of this type is typically a better fit for historical distributions than a normal one, especially as concerns fat tails.
        \end{enumerate}
        The plausibility of $\bm{S}_0$ can now be easily evaluated using just (\ref{eq::maha_def}). Assuming a normal distribution, the resulting value is compared to the quantiles of a $\chi^2(n)$ to determine the probability of a scenario equally or less extreme than $\bm{S}_0$. For other elliptical distributions where the law of the Mahalanobis distance is not known, a numerical solution exists\footnote{A more mathematical method also exists based on the density derivation formula for $Y=g(\bm{S})$ knowing the density $f$ of $\bm{S}$: $f_Y(t)=f\circ g^{-1}(t)\frac{dg^{-1}}{dt}(t)$. Here, $g$ corresponds to the Mahalanobis distance introduced in (\ref{eq::maha_def}). Though more interesting from a theoretical point of view, this solution may lead to long complex equations.}. First, the elliptical distribution that best fits $\bm{S}$ is determined. Second, a Monte Carlo simulation of $\bm{S}$ is performed. Then, approximate quantiles of the Mahalanobis distance are computed and probability $\alpha_{0,\text{approx}}$ of $\bm{S}_0$ can be deduced.
    \subsection{Fitting the Plausibility of a Given Scenario}
        If $\alpha_{0}$ or $\alpha_{0,\text{approx}}$ exceeds a given threshold $\alpha_{\max}$, then $\bm{S}_0$ lies outside of the admissible ellipsoid $\mathcal{E}^{\alpha_{\max}}$. In this case, the closest admissible scenario $\tilde{\bm{S}}$ to $\bm{S}_0$ on $\mathcal{E}^{\alpha_{\max}}$ is defined by homothety. This definition provides for minimal corrections in the terms of $\bm{S}_0$ thus conserving as much as possible the intuition of the portfolio manager.\par
        For the sake of clarity, the non-constraining assumption $\bm{\mu}=\bm{0}$ is made. Then:
        \begin{equation}
            \tilde{\bm{S}}=K\bm{S}_0,\ K\in\mathbb{R}
        \end{equation}
        As $\tilde{\bm{S}}\in\mathcal{E}^{\alpha_{\max}}$ it follows that $\tilde{\bm{S}}^\top\mathbf{\Sigma}^{-1}\tilde{\bm{S}}=q^{\alpha_{\max}}$ where $q^{\alpha_{\max}}$ is the $\alpha_{\max}$ quantile of the density of the squared Mahalanobis distance. This constraint leads to:
        \begin{equation}\label{eq:enhanced_scen}
            \tilde{\bm{S}}=\sqrt{q^{\alpha_{\max}}}\frac{\bm{S}_0}{\sqrt{\bm{S}_0^\top\mathbf{\Sigma}\bm{S}_0}}
        \end{equation}
    \subsection{Application}
        A portfolio manager runs two long/short strategies, each based on a different spread: the first on equity indices and the other on another asset class. The correlation $\rho$ between spreads is low: $\rho=0.25$. For the sake of argument, the spreads are assumed to have constant annual volatility $\sigma=10\%$.\par 
        The manager would like to know if, under these assumptions, the spreads scenarios have a strong probability (50\% for example) of incurring 10\% and 20\% losses over one year, leading to the scenario:
        \begin{equation}
            \bm{S}_0=[-10\%,-20\%]
        \end{equation}
        In this example, it is assumed that the spreads have either a normal or a Student's t with 2 degrees of freedom distribution. In reality however, the parameters of the elliptical distribution of reference would be determined using maximum likelihood estimators derived from the historical distribution. Thus, $\bm{S}_0$ corresponds to a 88.2\% (resp. 68.0\%) probability for a normal (resp. a Student's t) distribution. Therefore, the loss the manager had in mind is less plausible then expected. Setting $\alpha_{\max}=50\%$ in (\ref{eq:enhanced_scen}), the fitted scenario of interest for the manager is:
        \begin{align}
            \tilde{\bm{S}}&=[-7.5\%,-15.0\%]&\text{ for normal risk factors}\\
            &=[-5.8\%,-11.5\%]&\text{ for Student's t risk factors}\nonumber
        \end{align}
        Thus, the fitted scenarios respect the directions intended by the portfolio manager. Only the amplitude of the shocks is changed to comply with the scenario plausibility constraint.\par
        Obviously one can argue that the correlation and volatility used in that example do not reflect a crisis environment where $\bm{S}_0$ occurs. An extension of this example would therefore be to stress the correlation and volatility to best reflect a financial crisis environment. This process is further explained in section \ref{sec:stress_cor_mat}.
\section{Starting from Plausibility}\label{sec:second_approach} 
    The plausibility-driven ERST returns both the most extreme loss and a corresponding scenario for a given level of plausibility. This approach is studied in \cite{Studer} and further discussed in \cite{Breuer} and \cite{Breuer_2} for example. The advantage is that it returns a loss that may be compared to other existing risk measures such as VaR, which is briefly introduced in section \ref{sec:second_approach_VAR}. As shown in sections \ref{sec:second_approach_lin} from \ref{sec:second_approach_quad}, a plausibility-driven ERST is linearly dependent on VaR for linear and some non-linear portfolios. However, this relationship is not present as a general rule for non-linear portfolios, making plausibility-driven ERST interesting and valuable. For non-linear portfolios, the approach can be seen as a continuum of VaR and Expected Shortfall (ES) and sets a new paradigm for risk measurement. Some limitations do exist, however, as discussed in section \ref{sec:second_approach_shortcomings}.  
    \subsection{Existing Value-at-Risk Approach}\label{sec:second_approach_VAR}
        For a given $\alpha\in[0,1]$, VaR$_\alpha$ returns the $\alpha$ quantile of the P\&L density indicating that P\&L is not as extreme as the VaR output $\alpha\%$ of the time. The P\&L density may be historical or any other fitted density.\par
        Taking the simple case of linear portfolio with $n$ risk factors and weighting scheme $\bm{\omega}$, and assuming that risk factors are normally distributed $\bm{S}\sim\mathcal{N}(\bm{0},\mathbf{\Sigma})$, then P\&L$(\bm{S})\sim\mathcal{N}(0,\bm{\omega}^\top\mathbf{\Sigma}\bm{\omega})$ and: 
        \begin{equation}
            \text{VaR}_\alpha=-\mathcal{N}^{-1}(\alpha)\sqrt{\bm{\omega}^\top\mathbf{\Sigma}\bm{\omega}}\label{eq:VAR_lin}
        \end{equation}
        for $\mathcal{N}^{-1}(\alpha)$ the $\alpha$ quantile of a standard normal distribution.\par 
        For a quadratic portfolio, this expression does not hold true. The distribution of P\&L$(\bm{S})$ may not be analytically known depending on the density function for $\bm{S}$. However, an approximate VaR can be calculated after Monte Carlo simulations of $\bm{S}$ and derivation of the probability based on the resulting P\&L's distribution.\par
        With such an approach, VaR provides only a loss as output. This does not allow a portfolio manager to dig deeper and understand where the underlying weaknesses in  portfolio exposures lie.\footnote{This is possible with Historical VaR but only for historical / past scenarios.} In this respect, the plausibility-driven ERST provides a more complete result than VaR. In addition to a resulting loss, it provides a corresponding scenario, identifying specific strengths and weaknesses of the portfolio in order to take potentially countermeasures such as hedging or portfolio adjustments. This advantage is further detailed below.
    \subsection{Problem Statement}
        Let $\alpha$ and MaxERST\footnote{MaxERST was first introduced in \cite{Studer} and denoted as Maximum Loss (ML).} be the input level of plausibility and the output loss. The plausibility-driven ERST is then the optimization problem:
        \begin{equation}\label{eq:RST_second_approach}
            \min_{\text{Maha}^2(\bm{S})\leq q^{\alpha}}\text{P\&L}(\bm{S})
        \end{equation}
        In the two following sections, this problem is solved for both linear and quadratic portfolios. 
    \subsection{Application for Delta-One Strategies}\label{sec:second_approach_lin}   
        Assuming a long/short strategy on two momentum indices, $\text{P\&L}(\bm{S})=\bm{\omega}^\top\bm{S}$ with $\bm{\omega}=(1,-1)$. Here, (\ref{eq:RST_second_approach}) can be solved by relying on Lagrangian optimization with Kuhn-Tucker conditions. For a given $\alpha$, MaxERST and the corresponding scenario $\bm{S}^{\alpha}$ are:
        \begin{align}
            \bm{S}^{\alpha}=-\sqrt{q^\alpha}\frac{\mathbf{\Sigma}\bm{\omega}}{\sqrt{\bm{\omega}^\top\mathbf{\Sigma}\bm{\omega}}}\\
            \text{MaxERST}=-\sqrt{q^\alpha}\sqrt{\bm{\omega}^\top\mathbf{\Sigma}\bm{\omega}}\label{eq:RST_lin}
        \end{align}
        Comparing (\ref{eq:VAR_lin}) and (\ref{eq:RST_lin}), MaxERST and VaR are proportional. For linear portfolios, \cite{Breuer_3} states a similar relationship, adding that VaR and MaxERST are also proportional to the Expected Shortfall (ES) measure. The corresponding proof is in \cite{Sadefo}. Therefore when $\bm{S}$ is normally distributed:
        \begin{equation}\label{eq:VAR_RST_ratio}
            \frac{\text{VaR}}{\mathcal{N}^{-1}(\alpha)}=\frac{\text{MaxERST}}{\sqrt{q^\alpha}}=\frac{\text{ES}}{\rho(\alpha)\alpha}=-\sqrt{\bm{\omega}^\top\mathbf{\Sigma}\bm{\omega}}
        \end{equation}
        where $q^\alpha$ is the $\alpha$ quantile of a $\chi^2(n)$ distribution and $\rho(\alpha)$ the density of the standard normal distribution \cite{Breuer_3}. Despite being proportional for delta-one strategies, \cite{Breuer_3} argues that MaxERST is more useful than VAR. As it is sub-additive\footnote{i.e. absolute losses are such that $\text{MaxERST}(\text{portfolio 1})+\text{MaxERST}(\text{portfolio 2})\geq\text{MaxERST}(\text{portfolios 1 + 2})\geq 0$.} whereas VAR is not, MaxERST has proved to be a more reliable limit system than VAR for some simple non-linear portfolios such as some combinations of out-of-the-money short puts and short calls on the same underlying.\par
        For $q^\alpha$ the quantile of a $\chi^2(n)$ distribution, $\lim_{n\to\infty}q^\alpha=\infty$. Therefore, if $\bm{S}$ is normally distributed, the higher the number $n$ of risk factors the portfolio is exposed to, the more extreme MaxERST will be relative to VaR per (\ref{eq:VAR_RST_ratio}). This could create a dimensional dependency issue for irrelevant factors, as exposed in \cite{Mouy} or \cite{Breuer} and discussed in section \ref{sec:second_approach_shortcomings}.
        \begin{figure}
            \centering
            \begin{overpic}[width=.45\linewidth,trim={2.25cm 2.75cm 1.25cm 2.25cm},clip, frame]{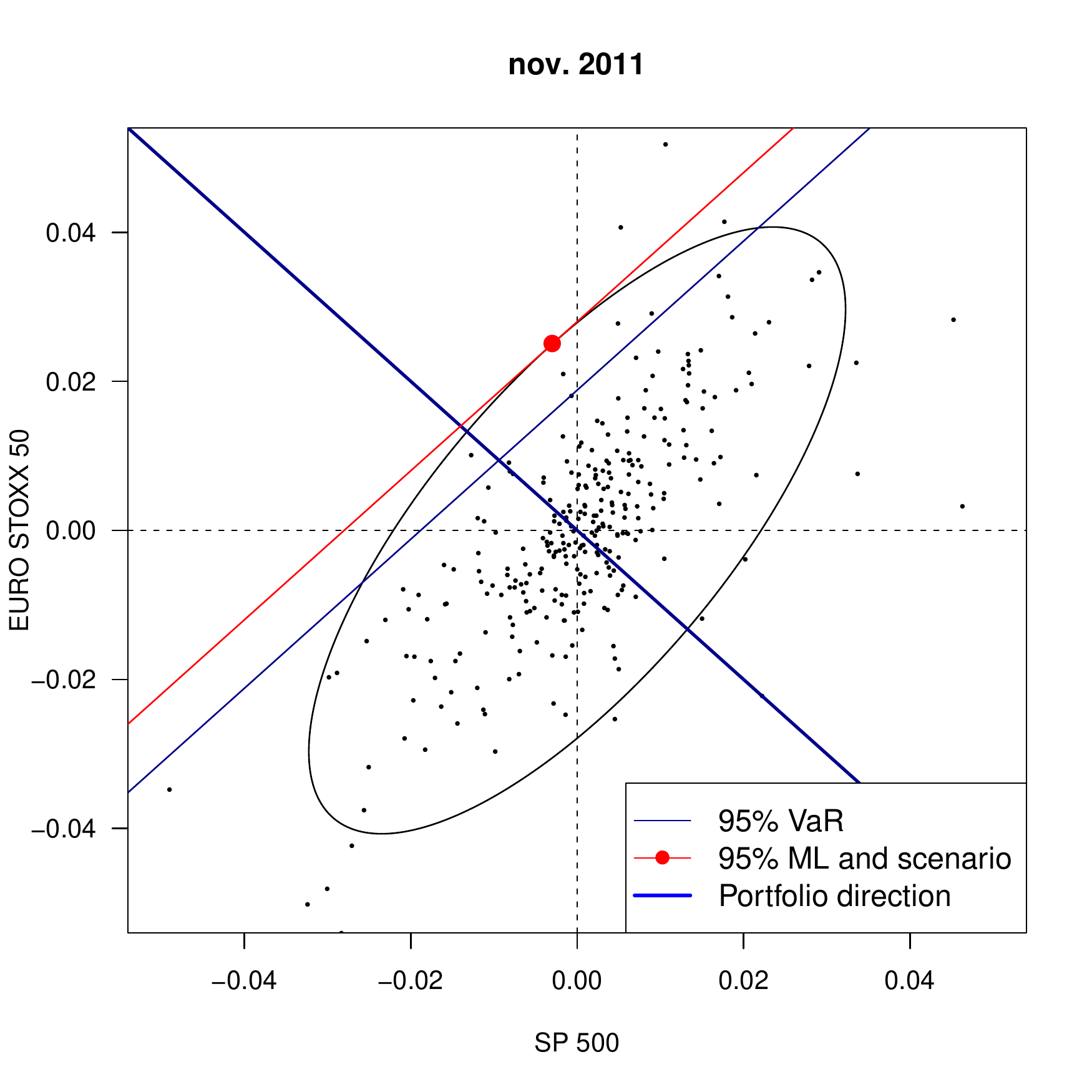}
                \put (93.5,2) {$\bm{\omega}$}
                \put (37.5,67.5){{\color{red}$\bm{S}^{95\%}$}}
            \end{overpic}
            \caption{Plausibility-driven ERST for a linear portfolio. 300 data points for two momentum indices (in grey) are used to compute $\mathbf{\Sigma}$. The level of plausibility is fixed at $\alpha=95\%$ and $\mathcal{E}^{95\%}$ is in black. Knowing the VaR and MaxERST per (\ref{eq:VAR_lin}) and (\ref{eq:RST_lin}), their corresponding Iso-P\&L lines are indicated.}
            \label{fig:longshort_maha}
        \end{figure}
    \subsection{Application for Non-Linear P\&L}\label{sec:second_approach_quad}
        For non-linear P\&L, a second order approximation is considered. Thus:
        \begin{equation}
            \label{eq:PnLexpression}
            \text{P\&L}(\bm{S})=\frac{1}{2}\bm{S}^\top\mathbf{A}\bm{S}+\mathbf{B}^\top\bm{S}
        \end{equation}
        where $\mathbf{A}$ and $\mathbf{B}$ are the respective second- and first-order sensitivities of the portfolio. Second-order sensitivities being symmetric, $\mathbf{A}$ is symmetric and, in the most general terms:
        \begin{equation}
            \mathbf{A}=
            \left[\begin{array}{c|c}
                \begin{matrix}
                    \ddots  &   & \pdv{\text{P\&L}}{S_i}{S_j}\\
                    & \pdv[2]{\text{P\&L}}{S_i} & \\
                    \pdv{\text{P\&L}}{S_j}{S_i} &   & \ddots           
                \end{matrix}
                & 
                \pdv{\text{P\&L}}{S_i}{\sigma_j}\\
                \hline
                \pdv{\text{P\&L}}{S_j}{\sigma_i}
                &
                \begin{matrix}
                    \ddots  &   & \pdv{\text{P\&L}}{\sigma_i}{\sigma_j}\\
                    & \pdv[2]{\text{P\&L}}{\sigma_i}    & \\
                    \pdv{\text{P\&L}}{\sigma_j}{\sigma_i}   &   & \ddots           
                \end{matrix}
            \end{array}\right]
            \hspace{2em}
            \mathbf{B}=
            \left[\begin{array}{c}
                \begin{matrix}
                    \vdots\\
                    \pdv{\text{P\&L}}{S_i}\\
                    \vdots\\
                    \hline
                    \vdots\\
                    \pdv{\text{P\&L}}{\sigma_i}\\
                    \vdots\\
                \end{matrix}
            \end{array} \right]
        \end{equation}\par
        With a quadratic form for P\&L, the resolution of (\ref{eq:RST_second_approach}) is more complex. The objective function may not be convex, therefore Kuhn-Tucker conditions are irrelevant. Fortunately, an optimization algorithm exists that can cope with this issue: the Levenberg-Marquardt algorithm. It is introduced in depth in \cite{Numerical_Optimization} and applied in \cite{Studer} to solve (\ref{eq:RST_second_approach}) with (\ref{eq:PnLexpression}).\par 
        For the same two momentum indices as in section \ref{sec:second_approach_lin}, results are shown in figure \ref{fig:quad_maha} for some $\mathbf{A}$ and $\mathbf{B}$.\par
        In addition, MaxERST is no longer linear with respect to VaR, as opposed to section \ref{sec:second_approach_lin}. This result justifies the use for ERST rather than VaR for non-linear portfolios as the approach is of added value for the portfolio manager and a continuum of VaR. Figure \ref{fig:VariableRhos} shows the aforementioned non-linearities. Yet, the specific case where $\mathbf{B}=\bm{0}$ remains linear as proved in Appendix \ref{ap:proof_guillaume}.
        \begin{figure}
            \centering
            \begin{overpic}[width=.45\linewidth,trim={2.25cm 2.75cm 1.25cm 2.25cm},clip, frame]{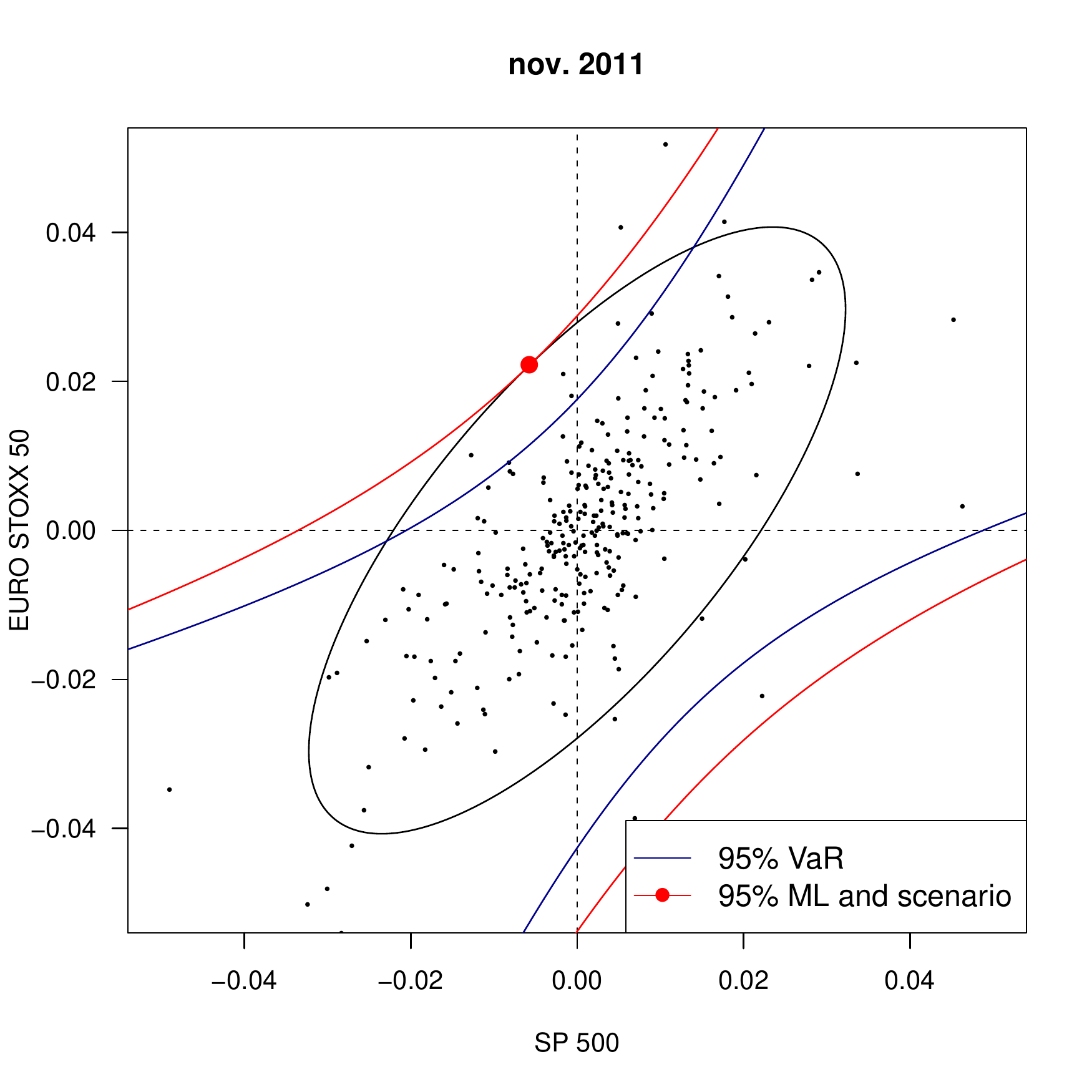}
                \put (35,65) {{\color{red}$\bm{S}^{95\%}$}}
            \end{overpic}
            \caption{Plausibility-driven ERST for a non-linear portfolio. Contrary to figure \ref{fig:longshort_maha}, the iso-P\&L lines are curved because of $\mathbf{A}$. The scenario $\bm{S}^{95\%}$ corresponding to the maximum loss within a $95\%$ plausibility constraint is found using the Levenberg-Marquardt optimization algorithm.}
            \label{fig:quad_maha}
        \end{figure}
        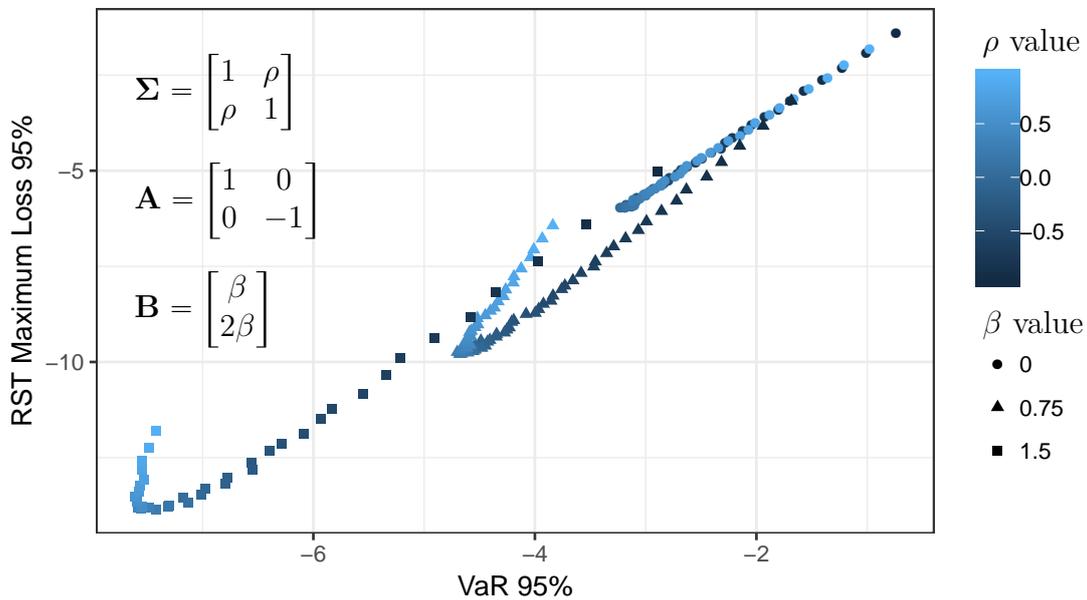
\begin{figure}
            \centering
            \begin{overpic}[width=.9\linewidth,clip]{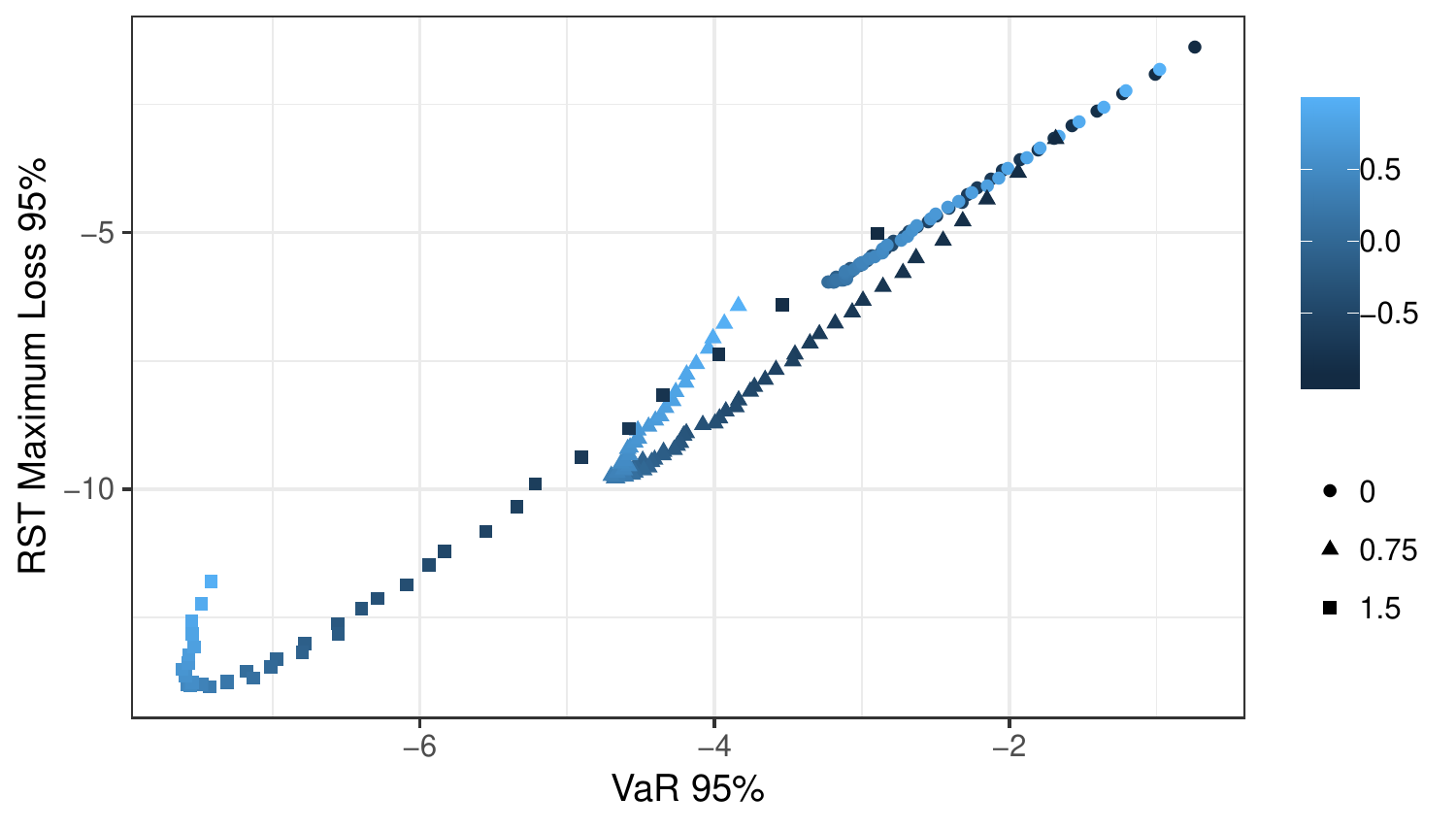}
                \put (12.5,47.5) {$\mathbf{\Sigma}=\left[\begin{matrix} 1 & \rho\\ \rho & 1\end{matrix}\right]$}
                \put (12.5,37.5) {$\mathbf{A}=\left[\begin{matrix} 1 & 0\\0 & -1\end{matrix}\right]$}
                \put (12.5,27.5){$\mathbf{B}=\left[\begin{matrix} \beta \\2\beta\end{matrix}\right]$}
                \put (90,26){$\beta$ value}
                \put (90,52){$\rho$ value}
            \end{overpic}
            \caption{Comparison between VaR and MaxERST outputs for $\alpha=95\%$ and two risk factors. The latter are simulated using unitary volatility for simplicity. For $\beta\neq 0$ the linear relation between VaR and MaxERST disappears. For $\beta=1.5$, VaR does not vary much with the positive correlation, whereas MaxERST does. The latter is therefore a more interesting risk measure here, as it reacts more strongly to moves in correlation.}
            \label{fig:VariableRhos}
        \end{figure}
    \subsection{On Dimensional Dependency}\label{sec:second_approach_shortcomings}
        As stressed in \cite{Mouy} and \cite{Breuer_2}, the output of the ERST approach depends directly on the dimension of the problem, i.e. the number $n$ of risk factors under consideration. Indeed, $q^\alpha$ in (\ref{eq:RST_second_approach}) varies with $n$. As previously stated, $q^\alpha$ is, for example, the quantile of a $\chi^2(n)$ distribution for a normally distributed $\bm{S}$.\par
        Therefore, adding risk factors to $\bm{S}$ even if they have a weight of zero in the portfolio does affect the output. Though this may be seen as a source of instability, this is also positive from a portfolio management perspective. It is actually a way to account for correlation with external yet meaningful risk factors that indirectly drive the variations. Nonetheless, it remains the portfolio and risk managers' responsibility to carefully select the external risk factors to consider.\par
        In addition, modifying the formulation of the problem resolves the dimensional dependency issue. Indeed, \cite{Rouvinez} overcomes this issue by replacing the quantile $q^\alpha$ in (\ref{eq:RST_second_approach}) by the plausibility of a given historical scenario $\bm{S}_H$ using (\ref{eq::maha_def}). This approach is more stable because the plausibility of $\bm{S}_H$ only varies with the risk factors that it actually depends on.  
\section{Starting from P\&L}\label{sec:third_approach}
    A P\&L-driven ERST extends the ideas expressed in \cite{Mouy} to non-linear portfolios. To this end, a new, adapted version of the Levenberg - Marquardt optimization algorithm is defined and tested. The main advantage of such an approach as compared to that in section \ref{sec:second_approach} is to overcome the problem stated in section \ref{sec:second_approach_shortcomings}.
    \subsection{Problem Statement}
        Given section \ref{sec:second_approach_shortcomings}, it is preferable that the constraint in (\ref{eq:RST_second_approach}) be independent of the squared Mahalanobis quantiles. Here, inverting the problem formulation works, i.e. finding the scenario with optimal plausibility for a given P\&L. This paves the way for the third and final approach discussed in this paper. The optimization problem becomes:
        \begin{equation}\label{eq:RST_third_approach}
            \min_{\text{P\&L}(\bm{S})= p}\text{Maha}^2(\bm{S})
        \end{equation}
        The case for a linear P\&L is discussed in \cite{Mouy}, but the resolution for non-linear P\&L remains outstanding. The remainder of this section focuses on this.
        \subsection{Resolution for Non-Linear P\&L}
        Rewriting (\ref{eq:RST_third_approach}) brings, for a loss $l$:
        \begin{equation}\label{eq:RST_quad}
            \min_{\frac{1}{2}\bm{S}^\top\mathbf{A}\bm{S}+\mathbf{B}^\top\bm{S}\leq l}\bm{S}^\top\mathbf{\Sigma}^{-1}\bm{S}
            =\min_{\frac{1}{2}\hat{\bm{S}}^\top\hat{\mathbf{A}}\hat{\bm{S}}+\hat{\mathbf{B}}^\top\hat{\bm{S}}\leq l}\Vert\hat{\bm{S}}\Vert^2
        \end{equation}
        where the change of variable $\hat{\bm{S}}=\bm{U}^{-\top}\bm{S}$ is performed with $\bm{U}$ the Cholesky decomposition matrix for $\mathbf{\Sigma}$ and:
        \begin{subequations}
            \begin{align}
                \hat{\mathbf{A}}&=\bm{U}\mathbf{A}\bm{U}^{\top}\\
                \hat{\mathbf{B}}&=\bm{U}\mathbf{B}
            \end{align}
        \end{subequations}
        Changing the variable allows the quadratic optimization problem to work with a centered bowl rather than an ellipsoid. The problem is thus reduced to finding the closest scenario(s)  $\hat{\bm{S}}^*$ to the origin and associated with the iso-loss curve of value $l$. This problem relates to the Levenberg-Marquardt optimization problem in section \ref{sec:second_approach}. However, the constraint is not necessarily convex here. Therefore, a new version of the method is introduced.\par
        It can be derived\footnote{This paper does not provide a rigorous proof of this statement, that would resemble the one in \cite{Numerical_Optimization} for a convex P\&L. For both clarity and applicability, the paper shows instead that the statement solves all the variations that optimization problem (\ref{eq:RST_third_approach}) takes.} from the equivalence proved in \cite{Numerical_Optimization}, Theorem 4.3, that $\hat{\bm{S}}^*$ is a solution to (\ref{eq:RST_quad}) if, and only if, it verifies the following conditions, for $\lambda_m$ the smallest eigenvalue of $\hat{\mathbf{A}}$ and for a given $\mu$:
        \begin{subequations}
            \begin{align}
                (\hat{\mathbf{A}}+\mu\mathbf{I})\hat{\bm{S}}^* &=-\hat{\mathbf{B}}\label{eq:cond1}\\
                \mu(\frac{1}{2}\hat{\bm{S}}^{*\top}\hat{\mathbf{A}}\hat{\bm{S}}^*+\hat{\mathbf{B}}^\top\hat{\bm{S}}^*-l) &=0\label{eq:cond2}\\
                \mu&\geq\max(0,-\lambda_m)\label{eq:cond3}
            \end{align}
        \end{subequations}
        The multidimensional optimization problem (\ref{eq:RST_quad}) reduces to a scalar optimization problem on $\mu$ under constraints (\ref{eq:cond1}) to (\ref{eq:cond3}). A problem of this type can be solved rapidly. As detailed below, a bisection algorithm to find the optimal $\mu$ allows for $\hat{\bm{S}}^*$ to be inferred directly.\par
        For $\mathcal{B}=(\bm{\pi}_1,\dots,\bm{\pi}_n)$ an orthonormal diagonalizing basis of symmetric matrix $\hat{\mathbf{A}}$:
        \begin{align}
            \hat{\bm{S}}^*=\sum_i\sigma_i\bm{\pi}_i\\
            \hat{\mathbf{B}}=\sum_i \beta_i\bm{\pi}_i
        \end{align}
        Defining $(\lambda_i)_i$ the eigenvalues of $\hat{\mathbf{A}}$, $I_m=\{i,\lambda_i=\lambda_m\}$ and taking (\ref{eq:cond3}) into account, (\ref{eq:cond1}) expressed in $\mathcal{B}$ becomes:
        \begin{subequations}
            \begin{align}
                &\sigma_i=-\frac{\beta_i}{\lambda_i+\mu}&\forall i\notin I_m\label{eq:cond_mu_1}\\
                &\sigma_j=-\frac{\beta_j}{\lambda_m+\mu}&\forall j\in I_m\text{ if }\mu\neq-\lambda_m\label{eq:cond_mu_2}\\
                &\sigma_j\in\mathbb{R}\text{ and }\beta_j=0&\forall j\in I_m\text{ if }\mu=-\lambda_m\label{eq:cond_mu_3}
            \end{align}
        \end{subequations}
        It is thus possible to find a unique $\hat{\bm{S}}^*$ if (\ref{eq:cond_mu_2}) is met and several $\hat{\bm{S}}^*$ are parametrized by $(\sigma_j)_{j\in I_m}$ if (\ref{eq:cond_mu_3}) is met. This result is important, because it illustrates the different cases of existence and unicity of $\hat{\bm{S}}^*$. In this respect, it is more complex than the functional expression obtained for the original Levenberg-Marquardt problem analyzed in \cite{Numerical_Optimization} and \cite{Studer}.\par
        Expressing (\ref{eq:cond2}) in $\mathcal{B}$ brings $\mu(f(\mu)-l)=0$ with:
        \begin{subequations}
            \begin{align}
                f(\mu)&=\sum_{i}\left[\frac{\lambda_i}{2}\left(\frac{\beta_i}{\lambda_i+\mu}\right)^2-\frac{\beta_i^2}{\lambda_i+\mu}\right]\text{ if }\mu\neq-\lambda_m\label{eq:PnL_mu_1}\\
                &=\sum_{i\notin I_m}\left[\frac{\lambda_i}{2}\left(\frac{\beta_i}{\lambda_i-\lambda_m}\right)^2-\frac{\beta_i^2}{\lambda_i-\lambda_m}\right]+\frac{\lambda_m}{2}\sum_{j\in I_m}\sigma_j^2\text{ if }\mu=-\lambda_m\label{eq:PnL_mu_2}
            \end{align}
        \end{subequations}
                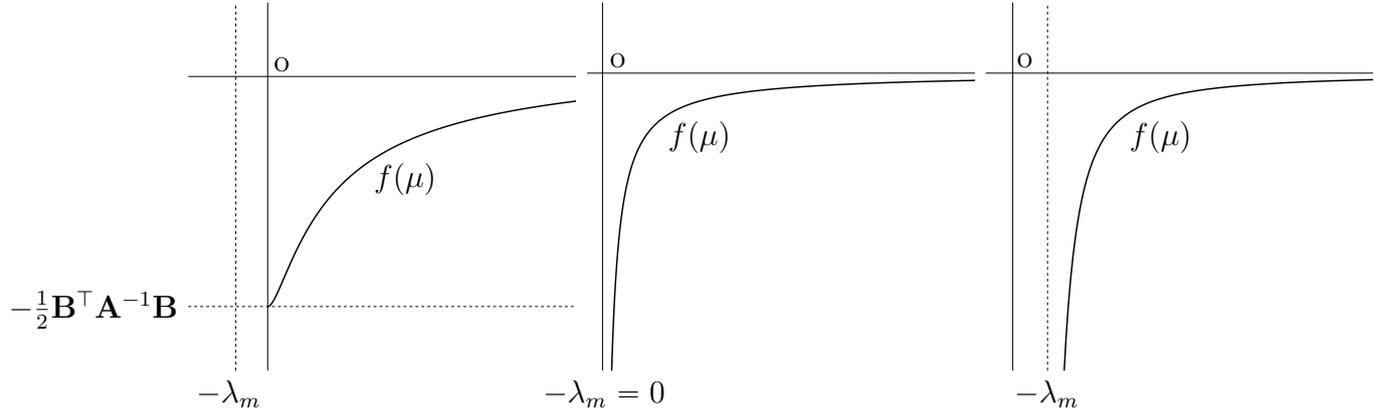
\begin{figure}
            \centering
            \begin{overpic}[width=.32\linewidth,trim={2cm 1cm 2cm 2cm},clip]{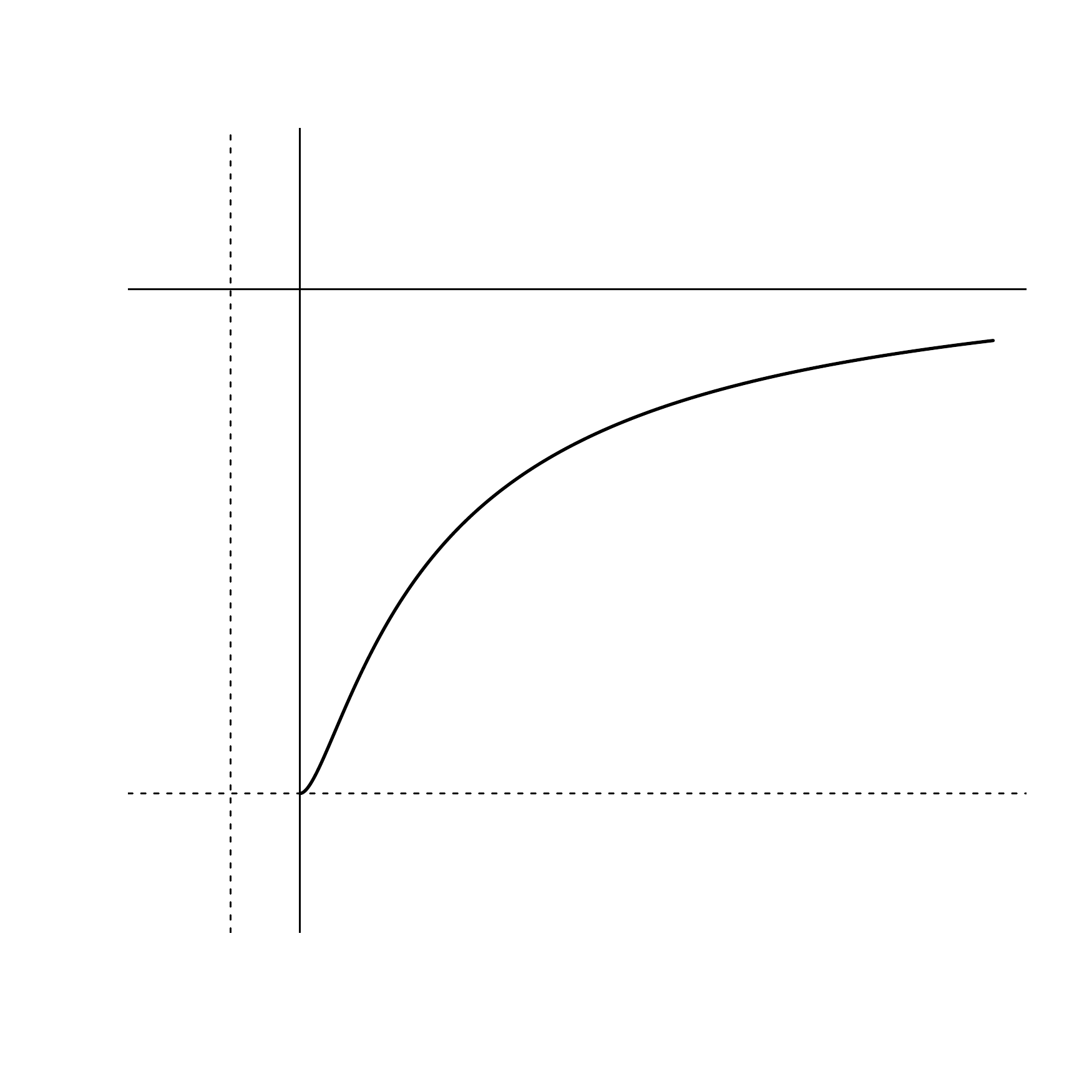}
                \put (21,83) {o}
                \put (45,55) {$f(\mu)$}
                \put (2.5,3) {$-\lambda_m$}
                \put (-42.5,23.5) {$-\frac{1}{2}\mathbf{B}^\top\mathbf{A}^{-1}\mathbf{B}$}
            \end{overpic}
            \begin{overpic}[width=.32\linewidth,trim={2cm 1cm 2cm 2cm},clip]{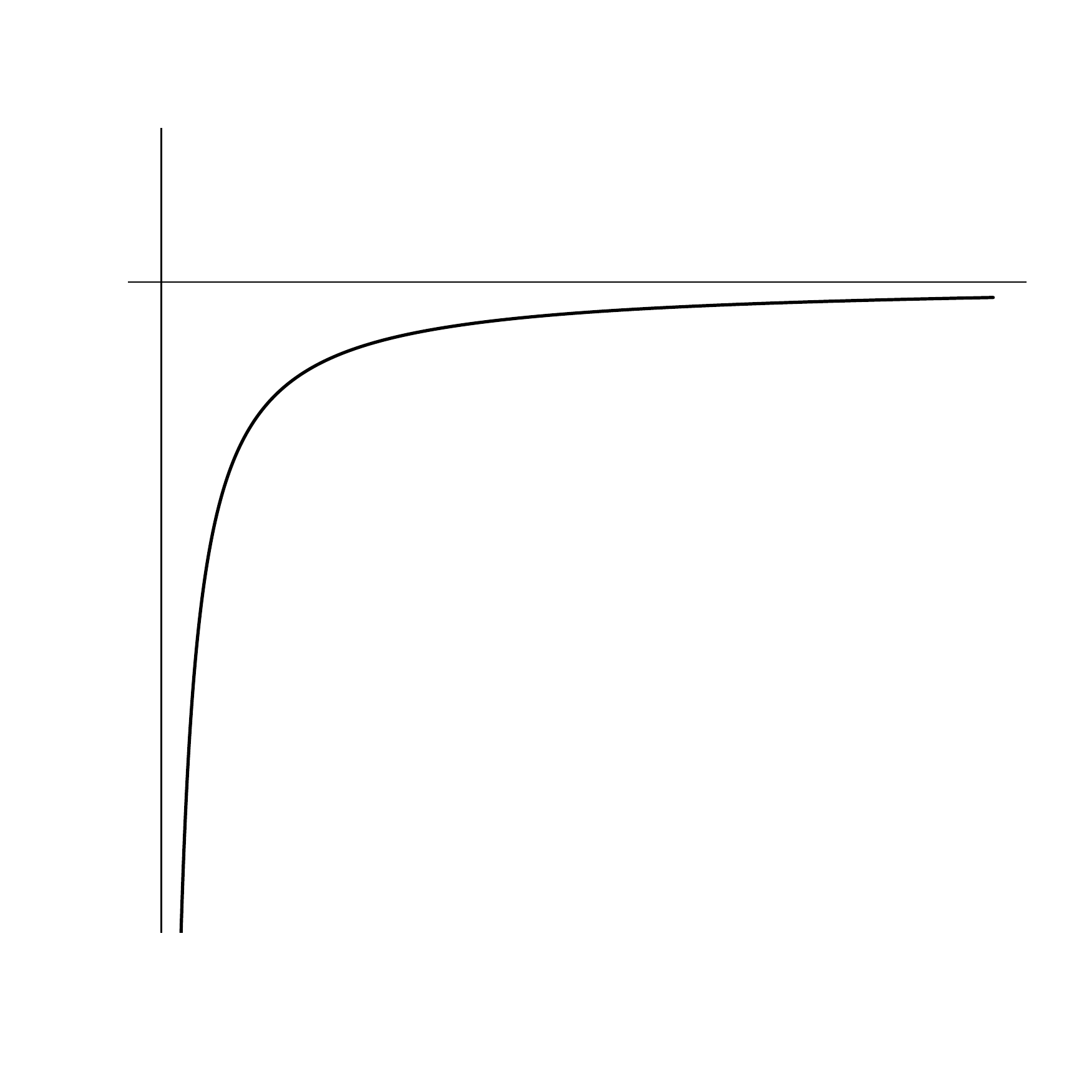}
                \put (6,84) {o}
                \put (20,65) {$f(\mu)$}
                \put (-10,3) {$-\lambda_m=0$}
            \end{overpic}
            \begin{overpic}[width=.32\linewidth,trim={2cm 1cm 2cm 2cm},clip]{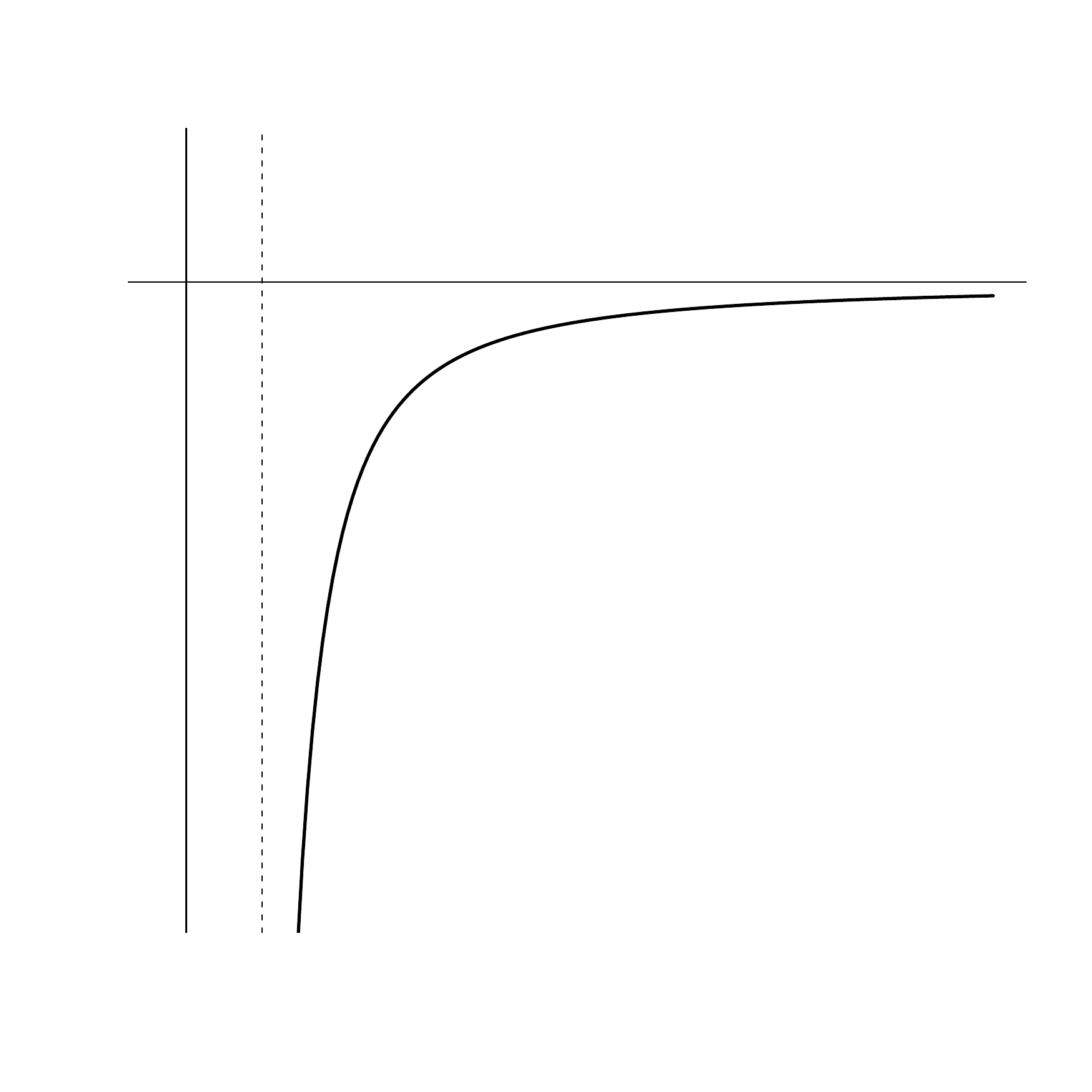}
                \put (8,84) {o}
                \put (35,65) {$f(\mu)$}
                \put (7.5,3) {$-\lambda_m$}
            \end{overpic}
            \caption{$f(\mu)$ when $-\lambda_m$ is negative, zero-valued or positive. This also means that $\hat{\mathbf{A}}$ is either strictly, semi or not positive-definite. The domain of definition for $\mu$ respects (\ref{eq:cond3}).}
            \label{fig:behaviours_f_mu}
        \end{figure}
        Different dynamics for $f$ are shown in figure \ref{fig:behaviours_f_mu}, which illustrates the following discussion on $\hat{\mathbf{A}}$:
        \begin{enumerate}
            \item For $\hat{\mathbf{A}}$ positive definite, $-\lambda_m<0$ and $\mu\geq 0$ per (\ref{eq:cond3}). 
                \begin{enumerate}
                    \item\label{item:1.a} If $\mu=0$,  $\hat{\bm{S}}^*=-\hat{\mathbf{A}}^{-1}\hat{\mathbf{B}}$ per (\ref{eq:cond1}) and $\text{P\&L}(\hat{\bm{S}}^*)=-\frac{1}{2}\hat{\mathbf{B}}^\top\hat{\mathbf{A}}^{-1}\hat{\mathbf{B}}=-\frac{1}{2}\mathbf{B}^\top\mathbf{A}^{-1}\mathbf{B}$, which corresponds to the global minimum P\&L. Such value for $\mu$ is chosen whenever the scenario for the global minimum is more plausible than the most plausible scenario for loss $l$.
                    \item\label{item:1.b} If $\mu\neq 0$, then the loss $l$ is attained per (\ref{eq:cond2}). However, such loss must be greater than the global minimum P\&L. $f$ being continuous and increasing, a single $\mu$ corresponds to any loss and can be approximated using a bisection algorithm.
                \end{enumerate} 
            \item For $\hat{\mathbf{A}}$ semi positive definite, $-\lambda_m=0$ and $\mu\geq 0$.
                \begin{enumerate}
                    \item\label{item:2.a} If $\mu=0$, (\ref{eq:cond_mu_3}) applies. Per (\ref{eq:PnL_mu_2}), the P\&L does not vary with any $\sigma_j$, $j\in I_m$ and the corresponding risk factors become irrelevant. The dimensions of the problem are thereby reduced and it becomes similar to \ref{item:1.a}. 
                    \item\label{item:2.b} If $\mu\neq 0$, then loss $l$ is attained per (\ref{eq:cond2}). As $\lim_{-\lambda_m^+}f=-\infty$  and $\lim_{+\infty}f=0$ and $f$ is still continuous and increasing, a single $\mu$ corresponds to any loss and can again be approximated using a bisection algorithm.
                \end{enumerate}
            \item For any other $\hat{\mathbf{A}}$, $-\lambda_m>0$ and $\mu\geq -\lambda_m$.
                \begin{enumerate}
                    \item\label{item:3.a} If $\mu=-\lambda_m$, (\ref{eq:cond_mu_3}) applies. Per (\ref{eq:PnL_mu_2}), the P\&L still varies with $\sigma_j$, $j\in I_m$. Constraint (\ref{eq:cond2}) becomes $f(\mu)=l$ and a root-finding algorithm (such as Newton Raphson) can determine which values $\sigma_j$, $j\in I_m$ must take. The solution may or may not be unique.   
                    \item\label{item:3.b} If $\mu\neq 0$, \ref{item:2.b} applies.
                \end{enumerate}
        \end{enumerate} 
       This indicates that it is only possible to solve (\ref{eq:RST_quad}) for losses ($l\leq 0$). This is actually a direct consequence of the formulation of the problem itself. Indeed, the null scenario always returns a zero-valued P\&L per (\ref{eq:PnLexpression}). In addition, the null scenario returns the lower boundary of the objective function in (\ref{eq:RST_quad}). Therefore, a profit input ($p>0$) cannot be obtained as a null scenario both returns a lower value in the objective function and respects the P\&L constraint. However, generating profit scenarios is of significant interest in assessing the asymmetries in portfolio P\&L. Thus, for $p>0$, (\ref{eq:RST_quad}) may be rewritten as follows:
        \begin{equation}
            \min_{-[\frac{1}{2}\hat{\bm{S}}^\top\hat{\mathbf{A}}\hat{\bm{S}}+\hat{\mathbf{B}}^\top\hat{\bm{S}}]\leq -p}\Vert\hat{\bm{S}}\Vert^2
        \end{equation}
    \subsection{Application to Non-Linear P\&L}
\begin{figure}
        \centering
        \includegraphics[width=.32\linewidth,trim={3cm 2.75cm 2cm 2.5cm},clip,frame]{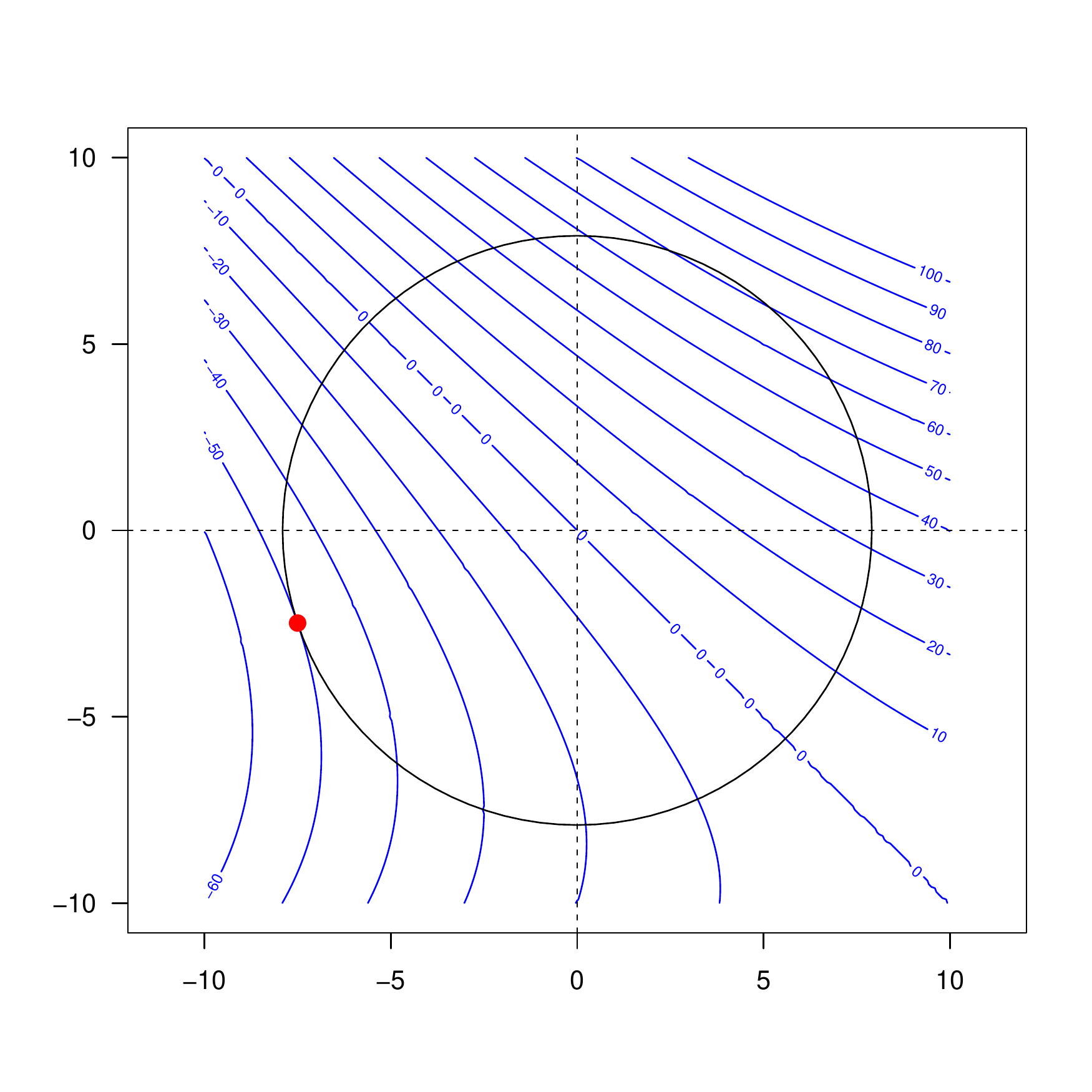}
        \includegraphics[width=.32\linewidth,trim={3cm 2.75cm 2cm 2.5cm},clip,frame]{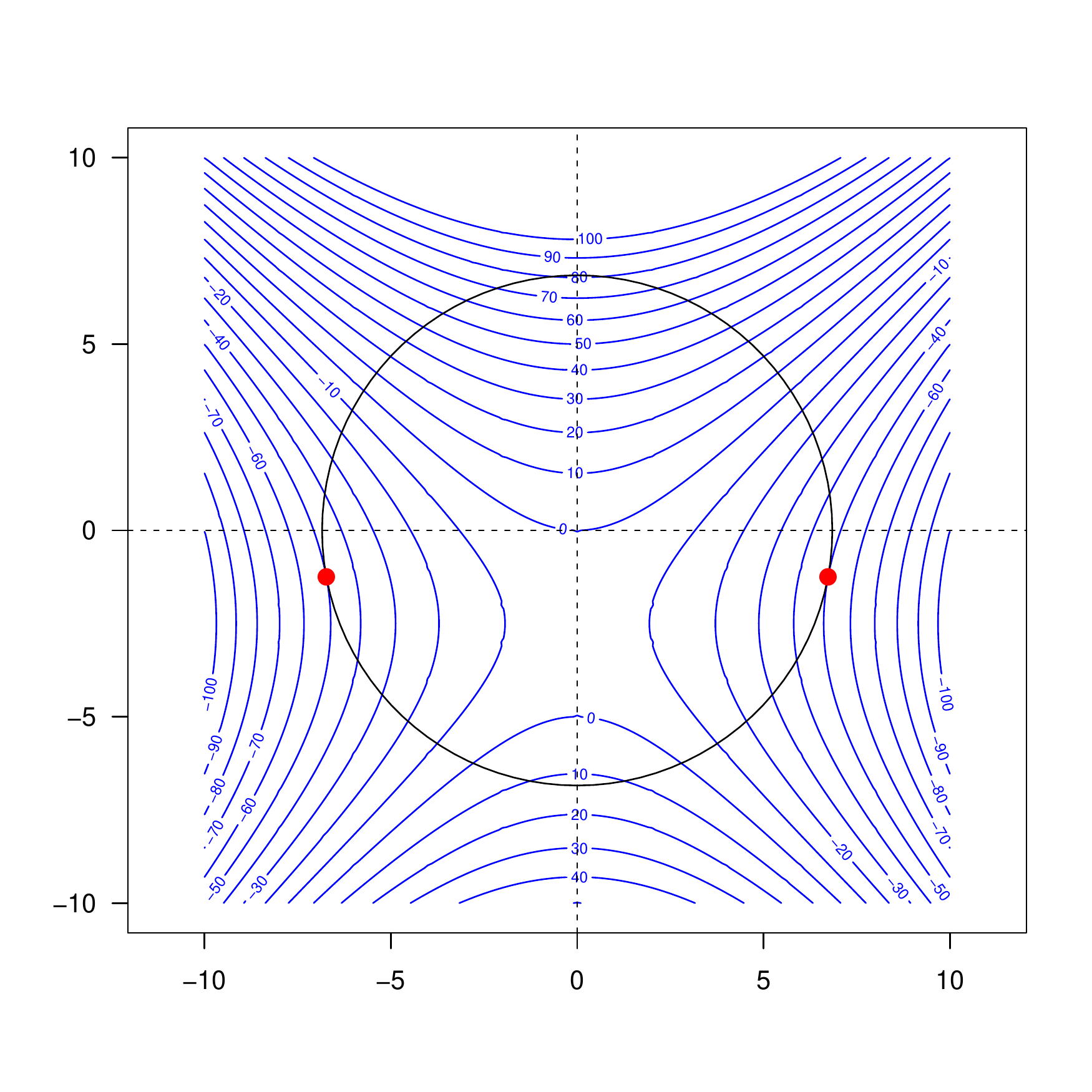}
        \includegraphics[width=.32\linewidth,trim={3cm 2.75cm 2cm 2.5cm},clip,frame]{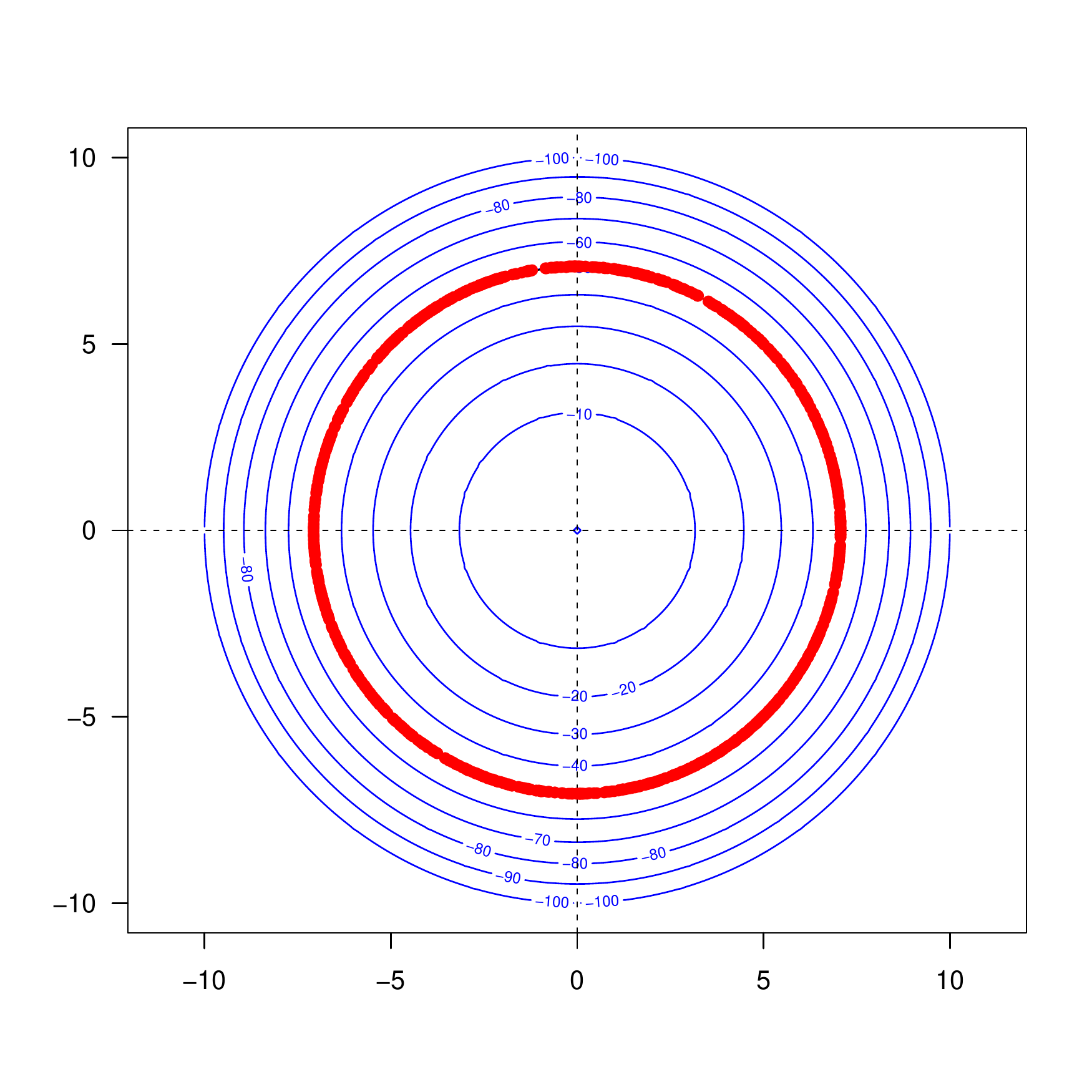}
        \caption{Solutions to problem (\ref{eq:RST_quad}) when $\hat{\bm{S}}$ consists of two risk factors. Depending on the P\&L expression, either one (left), two (middle) or an infinity (right) of solutions are found.}
        \label{fig:LM_adapted}
    \end{figure}

    The adapted Levenberg-Marquardt algorithm is tested on portfolios with two risk factors in figure \ref{fig:LM_adapted}.
\section{Additional Tools}
    Optional yet useful tools for all three approaches are briefly introduced in this section. They address either the challenges of reducing the dimension of the problem or the stress of the covariance matrix in (\ref{eq::maha_def}).
    \subsection{Reducing the Dimension}
        Many portfolios have a high number of risk factors. The above-mentioned quadratic optimization tools can still be run\footnote{As stressed in \cite{Studer}, the Levenberg Marquardt algorithm is time polynomial and therefore very efficient for nonconvex optimization problems.}, but the output becomes more difficult to interpret. To address this, two methods for reducing the number of risk factors are provided below. Of course, these methods imply a trade-off between the accuracy of the output and the amount of input data. 
        \subsubsection{Relying on the First Principal Component}
            The dimension can be reduced analytically using Principal Component Analysis (PCA). PCA quantifies how the variance-covariance matrix $\mathbf{\Sigma}$ of a multivariate random variable $\mathbf{X}$ responds to variations in each component $\left(X_i\right)_i$. PCA does this by determining the orthogonal directions (also known as Principal Components, or eigenvectors) to which $\mathbf{\Sigma}$ is exposed. The direction with maximum eigenvalue, or First Principal Component (FPC), is the one to which $\mathbf{\Sigma}$ is most exposed. The $\left(X_i\right)_i$'s with the most influence on $\mathbf{\Sigma}$ are those with highest coefficients in absolute value terms in the FPC. Indeed they lead the direction of the FPC.\par
            PCA would be performed on historical P\&L by risk factor, i.e. on series obtained from equation (\ref{eq:PnLexpression}) when applied separately to each risk factor. The $n$ most significant factors are then selected based on their coefficients in absolute value terms within the FPC and RST is performed on the reduced portfolio. 
        \subsubsection{Factor models}\label{sec:factor_models}
            If the FPC equally weights all risk factors, the previous method is no longer admissible. The variance of the portfolio is explained by all risk factors in equal proportion. In this case, grouping risk factors by cluster such as industry, market, country, index of reference, etc. is preferable to reduce the dimension of the problem and keep a satisfactory P\&L variance explanation.\par  
            Specifically, the beta of the risk factors in each cluster is calculated. Any of the RST approaches can then be performed directly on the clusters, thereby reducing the dimension. Finally, variations by risk factor can be reconstructed based on their beta values and the output of the RST.\par
            \cite{Packham} recently developed in more detail on factor models for RST with plausibility constraints for positively correlated risk factors. 
    \subsection{Stressing $\mathbf{\Sigma}$}\label{sec:stress_cor_mat}
        The plausibility of a scenario depends not only on the choice of the risk factor distribution (i.e. normal or Student's t distribution) but also on the covariance matrix $\mathbf{\Sigma}$ in (\ref{eq::maha_def}). If $\mathbf{\Sigma}$ corresponds to a specific historical stressed period characterized by both recorrelation and high volatility, the corresponding scenarios have low Mahalanobis distances. Inversely, if $\mathbf{\Sigma}$ corresponds to a calm period, the same scenarios have higher Mahalanobis distance and become less plausible. $\mathbf{\Sigma}$ therefore significantly impacts the RST output.\par
\begin{figure}
            \centering
            \begin{subfigure}[t]{.32\linewidth}
                \centering
                \includegraphics[width=\linewidth,trim={0cm 1cm 2cm 1cm},clip,frame]{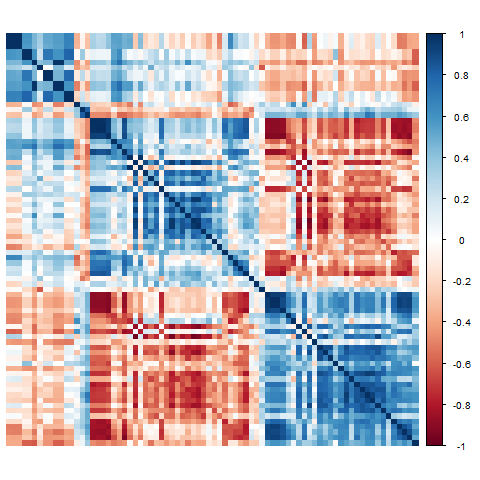}
                \caption{Historical February 2018 correlation matrix}
            \end{subfigure}
            \begin{subfigure}[t]{.32\linewidth}
                \centering
                \includegraphics[width=\linewidth,trim={0cm 1cm 2cm 1cm},clip,frame]{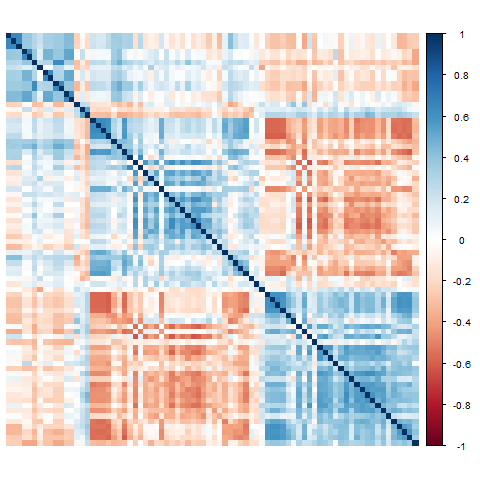}
                \caption{Shrinkage approximation}
            \end{subfigure}
            \begin{subfigure}[t]{.32\linewidth}
                \centering
                \includegraphics[width=\linewidth,trim={0cm 1cm 2cm 1cm},clip,frame]{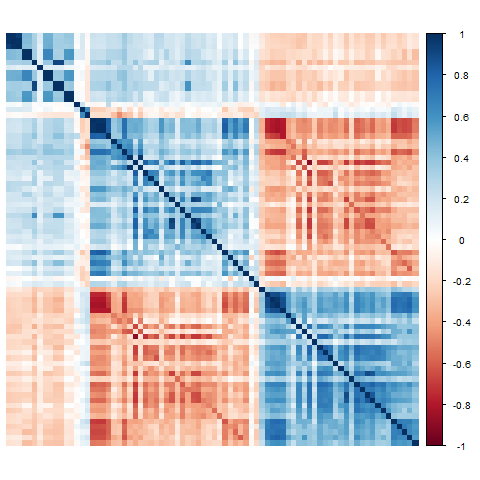}
                \caption{Single-factor stress model approximation starting from full year 2018 correlation matrix and $\theta = 0.25$.}
            \end{subfigure}
            \caption{the historical February 2018 correlation matrix (a) is not positive-definite if the number of risk factors exceeds the length of time series. However it can be approximated using positive definite matrices relying on shrinkage (b) or single-factor stress model (c). The three blocks of highly correlated data (in blue) consist of (from top to bottom): equity indices, single stocks and implied volatility. As expected, spot levels and volatility are negatively correlated (in red).}
            \label{fig:stress_cor_mat}
        \end{figure}

        Stressing $\mathbf{\Sigma}$ leads us to stressing the portfolio, which is of prior importance in risk management. Two methods are provided and illustrated in figure \ref{fig:stress_cor_mat} for the market crisis of February 2018 with a strong volatility spike in the US and to a lesser extent a short-lived fall in spot. This crisis saw recorrelation between supposedly independent premia strategies. Inverse de-correlation also occurred within long/short strategies: the short leg went up whereas the long leg went down for two historically positively correlated assets (for example, Euro-Stoxx 50 and S\&P 500 index implied volatility and spot levels on February 4, 2018).
        \subsubsection{Shrinking to a Short Period}
            This method derives an admissible $\mathbf{\Sigma}$ from a covariance matrix $\mathbf{\Sigma}_0$ computed on a short crisis period - typically from a week to a month.\par 
            Positive definiteness is required to compute the Mahalanobis distance - see (\ref{eq::maha_def}). Yet, $\mathbf{\Sigma}_0$ cannot be positive-definite if the number of risk factors exceeds the length of time series - see proof in Appendix \ref{Proof2}. Using longer windows of time would be counterproductive as changes in correlation average out and scenarios corresponding to a volatile environment are no longer plausible.\par 
            A preferable approach is to force the positive definiteness of $\mathbf{\Sigma}_0$ using shrinking methods, many of which are introduced and compared in \cite{Bun}. The advantage of shrinking is that it preserves the pattern of correlation and volatility observed during the crisis as much as possible. In figure \ref{fig:stress_cor_mat} for example, decorrelation (in red) within a highly correlated block of risk factors (in blue) is kept. Consequently, ERST approaches deem any scenario with similar dynamics to be likely which could lead to plausible scenarios being generated with stronger losses than those observed.
        \subsubsection{Single Factor Stress Model}
            This method stresses correlations within $\mathbf{\Sigma}$ where intensity is parametrized by $\theta$. This approach is introduced in \cite{JPM}. The purpose here is to increase negative and positive correlations block-wise and beyond the historical data.\par 
            Here, any historical period $[T_1,T_2]$ long enough for $\mathbf{\Sigma}$ to be positive definite is acceptable. Then, the series of the $n$ risk factors denoted by $(S_{i,t})$ with ${1\leq i\leq n}$ and ${T_1\leq t\leq T_2}$ is transformed as follows, for $0\leq\theta\leq 1$:
            \begin{equation}\label{eq:stressed_series}
                \hat{S}_{i,t}=(1-\theta)S_{i,t}+\frac{\theta}{n}\sum_{j=1}^{n}S_{j,t}
            \end{equation}
            The higher the $\theta$, the more correlated the series in (\ref{eq:stressed_series}) becomes as it is now a mixture of each previous series with a weight $(1-\theta)$ and the average of all previous series with a weight $\theta$. These correlations are then combined with the appropriate risk factor volatility to obtain the desired stressed version $\mathbf{\Sigma}_\theta$ of $\mathbf{\Sigma}$. As proved in \cite{JPM}, if $\mathbf{\Sigma}$ is positive definite, so is $\mathbf{\Sigma}_\theta$.\par
            A new refinement of this method is now available which allows for stressing of specific blocks within $\mathbf{\Sigma}$. This is of particular interest to stress the block corresponding to spot rates separately from that of volatility. For a block between rows and columns $n_1$ and $n_2$ of $\mathbf{\Sigma}$, (\ref{eq:stressed_series}) is rewritten as:
            \begin{equation}
                \hat{S}_{i,t}=(1-\theta^{\text{block}})S_{i,t}+\frac{\theta^{\text{block}}}{n_2-n_1+1}\sum_{j=n_1}^{n_2}S_{j,t}
            \end{equation}
            while the rest of the method remains unchanged.\par
            Finally, the value for $\theta$ may be derived to minimize the error $\varepsilon_\theta=\Vert\mathbf{\Sigma}-\mathbf{\Sigma}_\theta\Vert_{M}$ for a given matrix norm $\Vert.\Vert_{M}$. 
\section{Conclusion}
    The triptych approach of the Extended Reverse Stress Test (ERST) is defined in this paper, where with one input, either a scenario, a plausibility or a P\&L, the other two variables can be obtained. These methods are of added value as compared to traditional stress test and risk measures such as VaR or Expected Shortfall, mostly because the output contains more information. This additional information can help both portfolio and risk management teams to control a portfolio's sensitivities and re-allocate when needed.\par 
    To the authors' knowledge, several new developments are presented in this paper. In Section \ref{sec:first_approach}, applying a probability to a scenario a portfolio manager would like to test and suggesting a more plausible scenario are new. In Section \ref{sec:third_approach}, the adapted algorithm of the Levenberg-Marquardt optimization algorithm is new. In addition, section \ref{sec:second_approach} provides a complete view of the pros and cons of the most known and widely published approach of the three.\par
    Finally, ERST paves the way for a "new normal" in stress testing. In this respect, further research is needed. Possible next steps may include:
    \begin{itemize}
        \item A procedure for recomputing Greeks in (\ref{eq:PnLexpression}) to better account for their potential instability for scenarios with high market moves.
        \item A bootstrapping procedure for the covariance matrix $\mathbf{\Sigma}$ in (\ref{eq::maha_def}). This would mitigate the error in the estimated plausibility of a scenario due to the estimation of $\mathbf{\Sigma}$.
        \item A procedure for better interpreting any ERST output scenario. An interesting starting point may be the Maximum Loss Contribution defined in \cite{Breuer}.
        \item A procedure to go beyond the restriction of multivariate elliptical distributions. The use of copulas as in \cite{Mouy} would serve as a starting point.
        \item The development of a factor model able to model negative correlations for section \ref{sec:factor_models}.
    \end{itemize}
    \appendix
    \section{Purely Quadratic P\&L}\label{ap:proof_guillaume}
    For a portfolio with two risk factors $X$ and $Y$, covariance matrix is:
    \begin{equation*}
        \mathbf{\Sigma}=
        \begin{bmatrix}
            \sigma_X^2 & \rho \sigma_X \sigma_Y \\
            \rho \sigma_X \sigma_Y & \sigma_Y^2
        \end{bmatrix}
    \end{equation*}
    For $\sigma_X=\sigma_Y=\sigma$, eigenvalues $v_1$ and $v_2$ of $\mathbf{\Sigma}$ are $\sigma^2(1 \pm \rho)$ and corresponding eigen-vectors are:
    \begin{equation*}
        \mathbf{O}=[\bm{0}_1, \bm{0}_2]=\frac{1}{\sqrt{2}}
        \begin{bmatrix}
            1 & -1  \\
            1 & 1
        \end{bmatrix}
    \end{equation*}
    The P\&L is supposed to be purely quadratic, or pure Gamma i.e. $\text{P\&L}(\bm{S})= \bm{S}^\top\mathbf{A}\bm{S}$ with:
    \begin{equation*}
        \mathbf{A}=
        \begin{bmatrix}
            1 & 0  \\
            0 & -1
        \end{bmatrix}
    \end{equation*}
    Because of the negative contribution of the Gamma with respect to the y-axis, the minimum P\&L is reached on the frontier of the ellipsoid $\mathcal{E}^K=\{S | \bm{S}^\top\mathbf{\Sigma}^{-1}\bm{S} \leq K^2 \}$ for a given $K$. Such frontier can be represented as follows:
    \begin{equation*}
        \cos(\theta)\sqrt{v_1}K\bm{0}_1+\sin(\theta)\sqrt{v_2}K\bm{0}_2 = \frac{K}{\sqrt{2}}
        \begin{bmatrix}
            \cos(\theta)\sqrt{v_1} - \sin(\theta)\sqrt{v_2}\\
            \cos(\theta)\sqrt{v_1} + \sin(\theta)\sqrt{v_2}
        \end{bmatrix}
    \end{equation*}
    On $\mathcal{E}^K$, the P\&L can be rewritten as:
    \begin{equation*}
        \text{P\&L}(\theta)=-K^2\sigma\sqrt{1-\rho^2}\sin(2\theta)
    \end{equation*}
    For which the minimum is reached at $\theta=\frac{\pi}{4}$ and $\theta=\frac{5\pi}{4}$ with a value of $-K^2\sigma\sqrt{1-\rho^2}$.
    To comply with section \ref{sec:second_approach_quad}, risk factors are assumed to be normally distributed and $K^2=q^{\alpha}$ is the $\alpha$ quantile of a $\chi^2(2)$ distribution. Thus:
    \begin{equation}\label{eq:appRST}
        \text{MaxERST}=-q^{\alpha}\sigma\sqrt{1-\rho^2}
    \end{equation}
    VaR remains to be computed. From the projection of $\bm{S}$ on eigenvectors $\bm{0}_1$ and $\bm{0}_2$, it follows that $\bm{S}\sim \sqrt{v_1} X\bm{0}_1 + \sqrt{v_2}Y\bm{0}_2$ with $X$ and $Y$ independent standard normal random variables. Then:
    \begin{equation*}
        \text{P\&L}(\bm{S})=-2\sqrt{v_1v_2}XY
    \end{equation*}
    By independence, $\mathbb{E}[XY]=0$ and $\mathbb{V}[XY]=1$ and $XY$ can be approximated by a standard normal distribution. Thus:
    \begin{equation}\label{eq:appVAR}
        \text{VaR}=-2\mathcal{N}^{-1}(\alpha)\sigma \sqrt{1-\rho^2}
    \end{equation}
    From (\ref{eq:appRST}) and (\ref{eq:appVAR}) it follows that the plausibility-driven ERST and VaR approaches have linearly dependent output.
    
\section{Upper Bound of the Rank of a Covariance Matrix}\label{Proof2}
    For $\mathbf{X}$ the $n\times m$ historical matrix ($n$ days in rows) of $m$ series it follows that:
    \begin{equation*}
        \mathbf{\Sigma}=\left(\mathbf{X}-\overline{\mathbf{X}}\right)^\top\left(\mathbf{X}-\overline{\mathbf{X}}\right)
    \end{equation*}
    As $\rank\left(\mathbf{X}\mathbf{Y}\right)\leq\min\left(\rank(\mathbf{X}),\rank(\mathbf{Y})\right)$,
    \begin{equation*}
        \rank\mathbf{\Sigma}\leq\rank \mathbf{X}\leq \min(n,m)
    \end{equation*}
    Therefore $\mathbf{\Sigma}$ is an $m\times m$ matrix which rank cannot exceed $n$ in case $n<m$.
    \printbibliography
\end{document}